\documentclass[aps,10pt,reprint,superscriptaddress,showpacs]{revtex4-2}

\usepackage{xcolor}
\usepackage{physics}
\usepackage{amssymb}
\usepackage{graphicx}
\usepackage{mathtools}
\usepackage{amsmath}
\usepackage{hyperref}

\usepackage{amsfonts}

\hypersetup{
colorlinks = true,
urlcolor = blue,
linkcolor = blue,
citecolor = blue
}

\newcommand{\overbar}[1]{\mkern 1.5mu\overline{\mkern-1.5mu#1\mkern-1.5mu}\mkern 1.5mu}
\newcommand\numberthis{\addtocounter{equation}{1}\tag{\theequation}}

\graphicspath{{figures/}}

\definecolor{mypurp}{rgb}{0.35, 0, 0.7}
\definecolor{mygreen}{rgb}{0.35, .7, 0.0}

\newcommand{\leotp}[1]{\textcolor{purple}{#1}}

\newcommand{\TUM}{\affiliation{Department of Physics, Technical University of Munich, 85748 Garching, Germany}}
\newcommand{\MCQST}{\affiliation{Munich Center for Quantum Science and Technology (MCQST), Schellingstr. 4, 80799 M{\"u}nchen, Germany}}
\newcommand{\Nottingham}{\affiliation{School of Physics and Astronomy, University of Nottingham, Nottingham, NG7 2RD, UK}}
\newcommand{\CQNE}{\affiliation{Centre for the Mathematics and Theoretical Physics of Quantum Non-Equilibrium Systems, University of Nottingham, Nottingham, NG7 2RD, UK}}
\newcommand{\USH}{\affiliation{Department of Physics and Astronomy, University of California at Riverside, Riverside, California 92521, USA}}

\begin{document}
\author{Yu-Jie Liu} \TUM \MCQST
\author{Kirill Shtengel} \USH
\author{Adam Smith} \TUM \Nottingham \CQNE
\author{Frank Pollmann} \TUM \MCQST

\title{Methods for simulating string-net states and anyons on a digital quantum computer}

\begin{abstract}
Finding physical realizations of topologically ordered states in experimental settings, from condensed matter to artificial quantum systems, has been the main challenge en route to utilizing their unconventional properties.
We show how to realize a large class of topologically ordered states and simulate their quasiparticle excitations on a digital quantum computer. 
To achieve this we design a set of linear-depth quantum circuits to generate ground states of general string-net models together with unitary open string operators to simulate the creation and braiding of abelian and non-abelian anyons. We show that the abelian (non-abelian) unitary string operators can be implemented with a constant (linear) depth quantum circuit.
Our scheme allows us to directly probe characteristic topological properties, including topological entanglement entropy, braiding statistics, and fusion channels of anyons.
Moreover, this set of efficiently prepared topologically ordered states has potential applications in the development of fault-tolerant quantum computers.
  
\end{abstract}
\maketitle
\section{Introduction}

Starting with the discovery of the fractional quantum Hall (FQH) effect  \cite{tsui:1982}, a rich zoo of topologically ordered phases of matter has been discovered \cite{Wen1990a,wen:2017}. 
Topologically ordered phases do not fall into the conventional Landau symmetry breaking paradigm but are instead characterized by their long-range entanglement.
This long-range entanglement can in turn give rise to some unique physical properties such as the topological ground state degeneracy dependent on the boundary conditions of the system.
A particularly exotic feature of topologically ordered phases in two-dimensional (2D) systems is manifested by emergent anyonic excitations that exhibit fractional abelian or non-abelian braiding statistics. 
Despite extensive experimental efforts to detect anyons, it is only recently signatures of the abelian anyonic statistics have been demonstrated in $v = 1/3$ FQH states via collision and interferometry experiments \cite{bartolomei:2020,nakamura:2020}. 
Meanwhile, even though experimental signatures consistent with non-abelian anyonic statistics have been observed in FQH states at $\nu=5/2$ and $7/2$~\cite{Willett2009a,Willett2013b,Willett2019}, its definitive confirmation remains elusive thus far. 

While the search for
physical systems
hosting topological order and methods of its detection remains an active area of research, another route toward realizing topological states is
through their simulation in suitable quantum systems.  
To this end, exactly solvable models play a special role; they can serve as a test bed for calibrating the system in order to realize the established properties before venturing into the unknown. 
A quintessential example of soluble models of topological order is provided by 
string-net models~\cite{levin:2005}. 
These are lattice models the low-energy physics of which is described by a non-chiral doubled topological quantum field theory (TQFT)~\cite{Turaev1992,Freedman:2004}. 
String-nets also have an intimate connection to quantum computation: 
Abelian string-net models can be regarded as a family of quantum error-correcting codes~\cite{Pankovich:2018} and certain non-abelian string-net models can be used for universal quantum computation in relation to Turaev-Viro codes~\cite{koenig:2010}.

While the realization of such topologically ordered states in physical systems is extremely challenging due to their long-range entanglement,
recent advances in the fabrication of controlled quantum systems---including programmable superconducting qubits~\cite{supremacy:2019, Satzinger2021} and  Rydberg atomic arrays~\cite{semeghini:2021,ruben:2021,samajdar:2021}---offer promising platforms for their
realization.
In the \emph{noisy-intermediate-scale-quantum} (NISQ) era, quantum simulations are
intrinsically noisy and limited by the size~\cite{Preskill2018} and thus an efficient scheme with shallow quantum circuits is  highly desirable. 

In this work, we show that generic string-net (ground) states can be efficiently realized and manipulated on a digital quantum computer with shallow unitary circuits. This work generalizes and goes beyond the algorithm recently implemented experimentally in Ref~\cite{Satzinger2021}.
Our protocol allows a preparation of topologically ordered states with $\mathcal{O}(l)$-depth quantum circuits of local gates, where $l$ is the smaller of the width and the height of the system.
Moreover, it is possible to characterize the topological order of the prepared states by entanglement and braiding statistics measurements. 
The creation and manipulation of the localized anyonic quasiparticles in the abelian and non-abelian phases rely on string-like quantum circuits. 
The depth of the required circuits is constant for abelian anyons and scales linearly with the separation of the non-abelian anyons. 
Based on the anyonic braiding, we also show how the fusion of anyons can be determined by an efficient interferometry measurements. 
This completes a toolkit for simulating the underlying TQFT of the string-net model. 
As a result, the scheme can be viewed as an efficient mapping from the gate-based computation to an anyon-based computation~\cite{Freedman:2002}. 
Due to the inherent noise of NISQ devices, efficient algorithms become particularly important for obtaining reliable results. 
A special case of our protocol has already been used to realize  $\mathbb{Z}_2$ topological order on the 31-qubit Sycamore quantum processor~\cite{Satzinger2021}.

It is worth mentioning that other explicit unitary constructions for string-net states are known. 
Letting $L$ be the perimeter of the system, a depth-$\mathcal{O}(L\log L)$ quantum circuit can be derived from entanglement renormalization \cite{aguado:2008,konig:2009}. 
The isometric tensor network representation \cite{zaletel:2020} of the string-net states can be regarded as a local quantum circuit of depth $\mathcal{O}(L)$~\cite{Soejima2020, wei:2021,Slattery:2021}. 
A subclass of the string-net model, the quantum double model and its anyonic excitations~\cite{kitaev:2003} can be simulated without the presence of a background Hamiltonian by involving measurement operations~\cite{Aguado2008,Brennen2009}.
Additionally, there are alternative protocols for extracting anyon statistics based on wavefunction overlaps~\cite{zhang:2012}, or using defects and lattice deformation such as in~\cite{koenig:2010}.
The paper is structured as follows: in Sections~\ref{construction} and~\ref{sec:ds_construct} we describe the circuit construction on the abelian examples of toric code and double semion model before moving to the general case in Section~\ref{sec:general_sn}. 
In Sections~\ref{sec:tee}  and~\ref{anyonbraiding} we discuss the measurement of the topological entanglement entropy and anyonic statistics that characterize the topological order. 
We then proceed to describe the circuits for braiding anyons in Section~\ref{sec:string_op}  with abelian and non-abelian examples given. 
We conclude after Section~\ref{section:exp} by discussing the possible experimental realizations and applications.

\section{Toric Code}\label{construction}
\begin{figure}
\centering
\includegraphics{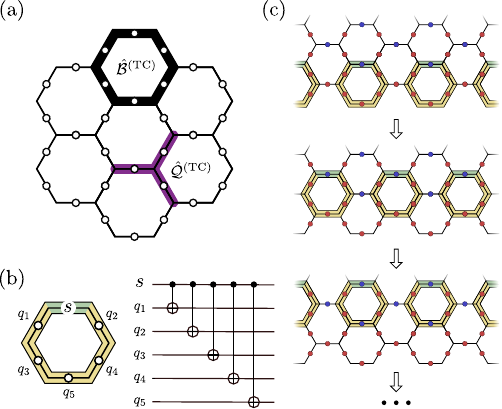}
\caption{The construction of the toric code ground state. (a) The honeycomb lattice with qubits on the edges that contains a  plaquette (black) and a vertex (purple); The vertices on the boundary only have two incoming legs. (b) The quantum gate C-$\hat{B}_p$ that is controlled by the representative qubit and targets the rest. The green marks the control qubits (unchanged upon C-$\hat B_p$), the yellow marks the target qubits (changed); (c) A graphical illustration of the algorithm. Prepare the red qubits in $\ket{0}$ and the blue qubits in $\ket{+} = (\ket{0}+\ket{1})/\sqrt{2}$. Then apply C-$\hat{B}_p$ on each plaquette in parallel for each row.}
\label{TC_construct}
\end{figure}

One of the simplest examples of topologically ordered states is the ground state of the toric code model (TC)~\cite{kitaev:2003}. This model consists of spin-$1/2$ degrees of freedom on the bonds of a honeycomb lattice with Hamiltonian
\begin{equation} \label{HTC}
    \hat{H}_{TC} = -\sum_s \hat{\mathcal{Q}}^{(\text{TC})}_s-\sum_{p} \hat{\mathcal{B}}^{(\text{TC})}_p.
\end{equation}
The commuting projectors $\hat{\mathcal{Q}}^{(\text{TC})}_s =\frac{1}{2}(1+ \prod_{j\in s}\hat{\sigma}^z_j)$ contain a product of Pauli matrices $\hat{\sigma}^z$ around each vertex $s$ and $\hat{\mathcal{B}}^{(\text{TC})}_p = \frac{1}{2}(1+\prod_{j\in p}\hat{\sigma}^x_j)$ contains a product of Pauli matrices $\hat{\sigma}^x$ around each plaquette $p$, as shown in Fig.~\ref{TC_construct}a.
Throughout this paper we consider open boundary conditions, as in Fig.~\ref{TC_construct}a, which are most relevant for experimental realization of the states on NISQ devices. 

In the Pauli-$z$ basis, we regard the qubit $\ket{1} = \ket{\downarrow}$ as being occupied by a string and  $\ket{0} = \ket{\uparrow}$ as unoccupied. The vertex and the plaquette projectors, $\hat{\mathcal{Q}}_s$ and $\hat{\mathcal{B}}_p$, constrain the ground state to be an equal-weight superposition of all the closed loop configurations. A ground state can be compactly written as a product of the plaquette projectors over the product state, neglecting the normalization:
\begin{equation} \label{eq:GSTC}
    \ket{GS}_{TC} \propto \prod_{p}\left(1+\prod_{j\in p}\hat{\sigma}_j^x\right)\ket{000\cdots 0}.
\end{equation}
With the choice of open boundary conditions, Eq.~\eqref{eq:GSTC} is the unique ground state of the Hamiltonian \eqref{HTC} \footnote{To be more precise, under this open boundary condition, some vertices on the boundary are only connected to two spins. The vertex operators $\hat Q$ on these vertices thus consist of Pauli matrices on two spins instead of three as in the bulk.}.
A similar idea can be adapted to periodic boundary conditions, where the ground state of the Hamiltonian becomes fourfold degenerate. 

The compact form of Eq.~\eqref{eq:GSTC} motivates an efficient quantum circuit construction for the ground state, which is as follows:
\begin{enumerate}
    \item Prepare an initial product state. We associate a representative qubit with each plaquette and they are initialized as $\ket{+} = (\ket{0}+\ket{1})/\sqrt{2}$ (blue in Fig.~\ref{TC_construct}c). The rest of the qubits are initialized as $\ket{0}$ (red in Fig.~\ref{TC_construct}c).
    \item
    Perform C-$\hat{B}_p$ over all the plaquettes in parallel on each row and iterate this from the bottom row to the top row. Where the controlled-$\hat B_p$ operation (C-$\hat B_p$ for short) applies CNOT gates controlled by the representative qubit to the rest of the qubits in the same plaquette, shown in Fig.~\ref{TC_construct}b.
\end{enumerate}
The depth of the quantum circuit built from this algorithm is linear in the number of plaquette rows in the system. Graphical illustrations of the steps are shown in Fig.~\ref{TC_construct}b-c. The C-$\hat{B}_p$ operator plays a central role in our construction of the quantum circuit. It flips all the rest of the qubits in the plaquette if the representative qubit is $\ket{1}$, otherwise it acts trivially. The  algorithm therefore has a nice interpretation that, each operation  C-$\hat B_p$, together with the initialization of the representative qubit for plaquette $p$, splits the state into an equal-weight superposition of being acted on by the identity or $\prod_{j\in p}\hat{\sigma}^x_j$. Iterating this set of operations over all the plaquettes yields the ground state of the form given in Eq.~\eqref{eq:GSTC}. If we follow this row-wise construction in Fig.~\ref{TC_construct}c and step 2, the circuit can be designed to have a depth that scales proportional to the smallest linear dimension of the system. A concrete experimental implementation of this algorithm can be found in Ref~\cite{Satzinger2021}.

The algorithm can be extended beyond the row-wise construction to other geometries, e.g. to periodic boundary conditions. However, the order must be such that each representative qubit is used as a control qubit before it is part of the targets of the C-$\hat{B}_p$ gate. That is, before applying the C-$\hat{B}_p$ gate, we require the representative qubit for that plaquette not to be entangled with the rest of the system. If the choice of representative qubits and the order in which the C-$\hat{B}_p$ gates are applied satisfy this constraint, then we say that the order is \emph{permissible}.
The row operation in step 2 of the algorithm can be replaced with other protocols following any permissible order. In general, the algorithm yields a parallel quantum circuit of depth that is at most linear in the perimeter of the system. We will see that this statement holds true for all the string-net models.   

\begin{figure}
    \centering
    \includegraphics{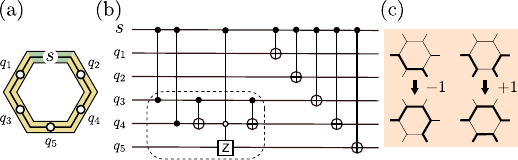}
    \caption{Construction of the double semion ground state. We implement the algorithm as in analogy to Fig.~\ref{TC_construct}c: the representative qubits (blue) are initialized in $\ket{-} = (\ket{0}-\ket{1})/\sqrt{2}$, and the rest of the qubits (red) are $\ket{0}$. (a) The construction consists of parallel application of C-$\hat{B}_p$ on each row. (b) The circuit diagram for C-$\hat{B}_p$ is shown. The two-qubit gate symbolised by two solid dots connected by a line is a controlled-Z gate. The three-qubit gate applies $\hat{\sigma}^z$ to the third qubit if the solid and hollow control qubits are 1 and 0, respectively, otherwise it does nothing. The dashed box in the circuit diagram contains the unitary that gives a phase +1(-1) if the number of loops changes (unchanged) when a TC plaquette operator is applied (as shown in (c)).}
    \label{DS_construct}
\end{figure}

\section{Double Semion Model}\label{sec:ds_construct}
Before moving on to  general string-net models, let us consider another spin-$1/2$ example---the double semion model (DS). This model can support semions and their chiral partners as excitations with an exchange phase of $\pm i$~\cite{Freedman:2004}. The Hamiltonian of the DS takes a form similar to Eq.~\eqref{HTC}, with $\hat{\mathcal{Q}}^{(\text{TC})}_s$ replaced by $\hat{\mathcal{Q}}^{(\text{DS})}_s$ and $\hat{\mathcal{B}}^{(\text{TC})}_p$ by $\hat{\mathcal{B}}^{(\text{DS})}_p$.
The vertex projector $\hat{\mathcal{Q}}^{(\text{DS})}_s$ projects onto the space where an even number of strings meet on each vertex $s$ (i.e. the same as the TC). The plaquette projector takes the form $\hat{\mathcal{B}}^{(\text{DS})}_p = \frac{1}{2}(1 - \hat{B}_p^1)$, where $\hat{B}_p^1$ flips all the qubits in the plaquette $p$ and associates a phase $+1(-1)$ to the configuration if the total number of loops in that configuration is changed (unchanged) after the flip. Note that in the literature, the factor of $-1$ in the DS model is typically associated with a change in the number of loops, but here we have an additional minus sign in front of $\hat{B}^1_p$, corresponding to the choice of $|-\rangle = (|0\rangle - |1\rangle) / \sqrt{2}$ for the representative qubits. This is only a matter of convention; we keep the minus sign here for later generalizations. 

Similarly to the TC, the resulting DS ground state can be obtained by applying a product of plaquette projectors $\hat{\mathcal{B}}^{(\text{DS})}_p$:
\begin{equation}\label{GSDS}
    \ket{GS}_{DS} \propto \prod_p \hat{\mathcal{B}}^{(\text{DS})}_p\ket{000\cdots0}.
\end{equation}
The DS ground state is a superposition of all the closed loop configurations with weights $(-1)^C$, where $C$ is the total number of closed loops in that configuration.

To construct the wavefunction given in Eq.~\eqref{GSDS}, we translate the same procedure from the TC construction: 
\begin{enumerate}
    \item We assign a representative qubit to each plaquette and initialize them in $\ket{-} = (|0\rangle - |1 \rangle)/\sqrt{2}$, the rest of the qubits are initialized in $\ket{0}$.
    \item Then apply the C-$\hat{B}_p$ operator over the plaquettes row by row in parallel as in Fig.~\ref{TC_construct}c.
\end{enumerate}
Notice the difference with the TC case: we initialize the representative qubits in the $\ket{-}$ state instead of $\ket{+}$. We also employ a different C-$\hat B_p$, which applies $\hat B^1_p$ to the other qubits in the plaquette if the representative qubit is $\ket{1}$ and acts trivially otherwise. An explicit circuit for the C-$\hat{B}_p$ is shown in Fig.~\ref{DS_construct}. Again during the construction, the order in which we apply C-$\hat{B}_p$ is important. We can choose the same row iteration as in the TC case, and then the depth of the circuits constructed row-wise scales linearly with the smallest linear dimension of the system. More generally we can follow any permissible order, as defined in Sec.~\ref{construction}. 

\begin{figure}
    \includegraphics{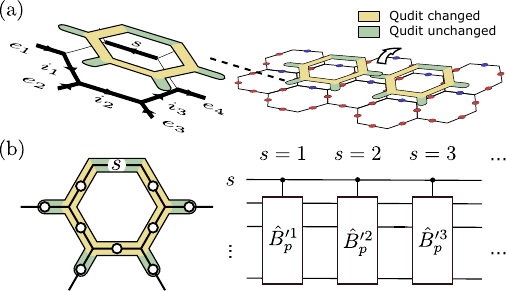}
    \caption{The construction for general string-net models. (a) During the construction, C-$\hat{B}_p^s$ only acts on the subspace shown on the left, here $i_1,i_2,i_3,e_1,e_2,e_3,e_4$ labels all the allowed strings under branching rules. $s$ is the state of the representative qudit. The arrows indicate the convention for the string orientation. One can prepare general string-net states on honeycomb lattice by applying C-$\hat{B}_p$ row by row. Despite the overlap of the gates, they can be decomposed for parallel implementation on each row. The C-$\hat{B}_p$ gate is controlled by the green qudits and alters the yellow ones. The representative qudits (blue) are initialized in $\mathcal{D}\sum_i a_i\ket{i}$. (b) The gate structure of C-$\hat{B}_p$; If the representative qudit is in state $\ket{s}$, it applies $\hat{B}'^s_p$ to the rest of the spins within the plaquette. This gate satisfies Eq.~\eqref{c_B}. We provide an explicit circuit for the gate in Appendix~\ref{append:explicitcircuit}}
    \label{general_sn}
\end{figure}

\section{General String-net Model}\label{sec:general_sn}
A general string-net model can be defined on a honeycomb (or any trivalent) lattice with local spins located on the edges. These spin degrees of freedom correspond to different {\it string types} at that edge. The strings are oriented in general, we use $i^*$ to denote the string $i$ with inverted orientation, with $i^*\neq i$ for oriented strings. String-net models describe a large class of topologically ordered phases the low-energy physics of which gives rise to a doubled TQFT~\cite{levin:2005}. On a quantum computing platform, the spins (or strings of different types) are encoded as {\it qudits}, a generalization of qubits with more than two levels, which in practice could correspond to multiple physical qubits. 

A string-net model is specified by a set of branching rules and self-consistent local constraints. The branching rules are all the triplets of string types $\{a,b,c\}$ that are allowed to meet at each trivalent vertex while open-ended strings are prohibited. For example, in the TC and DS model, the branching rules only allow qubits that add up to 0 (mod 2) to meet at the vertex. On the other hand, the local constraints ensure that any string configurations that can be smoothly deformed to each other have the same weight in the ground state. As a result, the system describes a fixed point of the real-space renormalization group flow \cite{konig:2009,gu:2009}. We give a brief summary of the general string-net model in Appendix~\ref{append:stringnetsummary}.

For a model with $N$ nontrivial string types, we can define an orthonormal basis $\{\ket{s}\}$, i.e. $(N+1)$-level qudit states, for string $s = 0,1\dots N$ where each qudit state $\ket{s}$ labels an edge occupied by a string $s$. The label $s = 0$ is reserved for the vacuum or null string. The Hamiltonian of the general string-net model generalizes the form of TC and DS, consisting of commuting projectors
\begin{equation}\label{H}
     \hat{H} = -\sum_v \hat{\mathcal{Q}}_v-\sum_{p}\hat{\mathcal{B}}_p,
\end{equation}
where $\hat{\mathcal{Q}}_v$  projects the triplet of strings at each vertex $v$ onto the allowed branching, $\hat{\mathcal{B}}_p = \sum_s a_s\hat{B}^s_p$ is the plaquette projector with $a_s$ being some real coefficients determined by specific string-net models. $\hat B_p^s$ is a plaquette operator for each string type $s$, it acts on the plaquette $p$ and its six external legs (see Appendix~\ref{append:stringnetsummary} for a detailed definition). By construction all the terms commute with each other. 

Notice that the TC and DS in the previous sections are special cases of Eq.~\eqref{H} with $a_0 = a_1 = 1/2$ and $a_0 =-a_1 = 1/2$ respectively. The plaquette operators are $\hat B_p^0 = 1$ and  the non-trivial plaquette operator $\hat B^1_p$ in each case. The coefficients $a_s$ are related to the {\it total quantum dimension} $\mathcal{D}$ of the underlying anyonic theory through $\mathcal{D} = 1/\sum_i a_s^2$. 

A ground state of the general string-net model can be conveniently written as a product of projectors over the zero product state
\begin{equation}
    \ket{GS} \propto \prod_p \hat{\mathcal{B}}_p\ket{000\cdots0} =\prod_p \left(\sum_s a_s\hat{B}^s_p\right)\ket{000\cdots0}.
\end{equation}
The state preparation is a direct generalization of the case for the DS, with the C-$\hat{B}_p$ (DS) operator acting on qubits replaced by the new C-$\hat{B}_p$ operator acting on qudits. Similar to controlled operations on qubits, this controlled operation applies a unitary to the target qudits if the control qudit is in a particular state $s$, as shown in Fig.~\ref{general_sn}.

For simplicity, we focus on the same row-wise construction as in the previous sections. A generalization to any permissible order is straightforward.
When C-$\hat{B}_p$ operations are performed from the bottom to the top row during the row-wise construction, the qudit subspace on which the qudit gates act is shown in Fig.~\ref{general_sn}a. The shape of general C-$\hat B_p$ generalizes those for the TC and DS; there are four additional qudits covered by the gates and the states of these qudits are unchanged upon the application of the gate. Despite the shared bond between the neighbouring operations, similar to the CNOT decomposition for the TC construction in Fig.~\ref{TC_construct}c, the C-$\hat{B}_p$ gate can be decomposed into a sequence of smaller controlled qudit gates around a plaquette. This allows for parallel implementation of the gates to all the plaquettes along each row (see Appendix~\ref{append:explicitcircuit}).

We can therefore adopt the following steps to construct a general string-net ground state:
\begin{enumerate}
    \item We assign a representative spin for each plaquette and initialize them in a state $\mathcal{\sqrt{D}}\sum_s a_i\ket{s}$. The rest of the spins are initialized to be $\ket{0}$.
    \item  We then iteratively apply the C-$\hat{B}_p$ on each row of plaquettes.
\end{enumerate}
Here operators C-$\hat{B}_p$ for the row construction satisfy
\begin{equation}\label{c_B}
   \includegraphics[scale = 1]{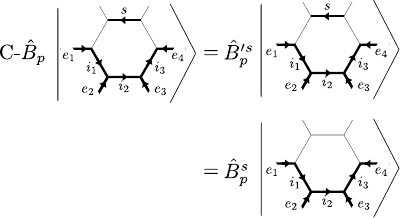}.
\end{equation}
 Thin unlabeled edges stand for the vacuum $\ket{0}$, while the set of labels $\{i\}$ and $\{e\}$ are arbitrary qudit configurations satisfying the branching rules of the string-net ground state. Edges with arrows indicate a convention that we use to label an oriented string with the qudit state. Operator $\hat B'^s_p$ is implicitly defined by Eq.~\ref{c_B} and takes into account that the representative qubit starts in the state $s$, such that its action corresponds to applying plaquette operator $\hat B^s_p$ to that plaquette. We indicate the strings with a particular convention for the orientation. However, for models with only unoriented strings, no convention needs to be chosen. This includes all the spin-$1/2$ string-net models, the TC, the DS, and the double Fibonacci model discussed in Section~\ref{section:exp}.
The C-$\hat B_p$ operation is controlled by the representative qudit: if the qudit is $\ket{s}$, $\hat{B}'^s_p$ is applied to the rest of the qudits in the same plaquette $p$ (see Fig.~\ref{general_sn}b), which is the same as acting with $\hat B_p^s$ on the state with a trivial representative qudit. 
It is also possible to implement the C-$\hat B_p$ operations in a different permissible order other than the row-wise construction above. To do this, a more generic definition of C-$\hat B_p$ is needed whereby the subspace of the initial qudit configurations is more general than that shown in Fig.~\ref{general_sn}a and Eq.~\eqref{c_B}. 
In Appendix~\ref{append:isometry}, we give a more general subspace of qudit configurations for defining C-$\hat B_p$ that can be used for any permissible order.
The operator C-$\hat B_p$ is well defined due to the isometry condition of $\hat{B}_p^s$ restricting to the subspace where at least one of the qudit states on the plaquette edge is trivial $\ket{0}$. A proof of the isometry property is given in Appendix~\ref{append:isometry}.

The construction prepares any given string-net ground states from $\mathcal{O}(l)$ layers of parallel local quantum gates, where $l$ is the smallest linear dimension of the system. The circuit depth shows explicitly that the lower bound of the circuit scaling provided in Ref.~\cite{bravyi:2006} is optimal for string-net states.

\begin{figure}
    \centering
    \includegraphics{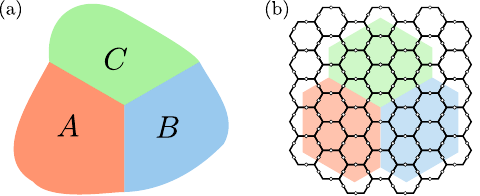}
    \caption{Probing the TEE on a quantum computer. (a) The simply connected domains $A,B$ and $C$ are used in the subtraction procedure given in Eq.~\eqref{stopo}; the exterior is the rest of the system. (b) A connected domain on a honeycomb lattice for extracting the TEE. The domain includes the qudits on the external legs, while the qudits on the boundary of two domains are shared by both sides. }
    \label{tee}
\end{figure}

\section{Topological Entanglement Entropy}\label{sec:tee}
Topological order is characterized by the presence of long-range entanglement. Measurement of the long-range entanglement of string-net states allows us to verify and partly characterize the topological order of the string-net states that we can efficiently prepare. Such entropy measurements are generally costly but are still feasible on small-scale experiments on current and near-term devices~\cite{brydges:2019,elben:2019, Huang:2020}, as experimentally demonstrated in Ref.~\cite{Satzinger2021}. For the sake of completeness, we recall the main idea of this entanglement measure and its detection on honeycomb lattice.

The entanglement entropy of subsystem $A$ is defined as the von Neumann entropy $S_A = -\text{tr}(\rho_A\ln\rho_A)$, where $\rho_A$ is the reduced density matrix of the subsystem $A$. If we consider a disk with boundary length $L$ and compute the entanglement entropy for the region enclosed by the disk, it is expected to follow $S = \alpha L -\gamma + \cdots$, where $\alpha$ and $\gamma$ are constants. The rest of the terms vanish as $L\to\infty$. This so-called {\it area law} has been proven for 1D gapped Hamiltonian~\cite{Hastings:2007} and conjectured for higher dimensions.

The constant $\alpha$ is non-universal and depends on the details of the Hamiltonian, but $\gamma(\geq 0)$ is a universal quantity that cannot be changed unless a quantum phase transition occurs. The negativity of $-\gamma$ and the positivity of von Neumann entropy prevent one from smoothly connecting the state to a product state without changing $\gamma$ (via a quantum phase transition), indicating an intrinsic long-range entanglement. This universal quantity  $S_\text{topo} = -\gamma$ is dubbed the {\it topological entanglement entropy} (TEE)  \cite{kitaev:2006, Levin2006} and can be determined in a model-independent way via a subtraction procedure 
\begin{equation}\label{stopo}
    S_{\text{topo}} = S_A +S_B+S_C-S_{AB}-S_{BC}-S_{AC}+S_{ABC},
\end{equation}
where $A, B$ and $C$ are simply connected domains (Fig.~\ref{tee}a) and they are assumed to be large compared to the correlation length. The TEE is an invariant in the sense that it does not change when $A,B,C$ are smoothly deformed. It therefore serves as a useful probe to characterize a topological order of the system.

The string-net ground states are known to obey the area law and have $S_\text{topo} = -\log\mathcal{D}$, where $\mathcal{D}$ is the total quantum dimension. In a model with abelian anyons, the quantum dimension simply counts the total number of anyon species $\mathcal{D} = \sqrt{\text{\# anyons}}$, which for the TC and DS give $S_\text{topo} = -\ln 2$. The zero correlation length in string-net states allows for accurate inference of the TEE from very small subsystems and ensures that the TEE stays invariant even when the region outside $ABC$ is truncated away in the state construction.

We can extract the TEE from the system by performing a full quantum state tomography (QST). This is generally impractical for large system size as it involves reconstructing the reduced density matrices by measuring a complete set of observables. 
A nice feature of the string-net states is that their TEE remains unchanged if the von Neumann entropy is replaced by the R\'enyi entropy of arbitrary order \cite{flammia:2009}. This provides an alternative approach to accessing TEE via randomized measurements \cite{elben:2019, Huang:2020}. The protocol allows for a direct access to the entropy without reconstructing the state, which significantly reduces the necessary measurements needed for statistical averaging. When operating on NISQ devices, the randomized measurements can be performed together with the state-of-art error-mitigation technique to produce an unbiased estimator for the TEE, as successfully demonstrated for subsystems up to nine qubits in Ref.~\cite{Satzinger2021}.

\begin{figure}
    \centering
    \includegraphics{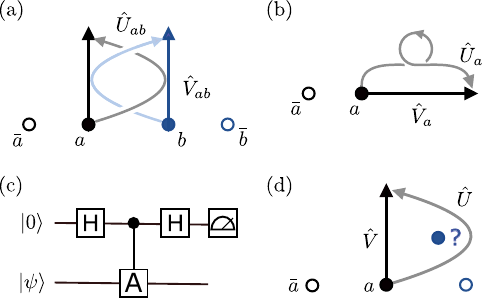}
    \caption{Probing braiding statistics by interferometry. The unitary paths $\hat V$ and $\hat U$ for obtaining (a) $S_{ab}$ and (b) the twist factor $\theta_a$.  (c) A simple circuit for measuring the expectation $\langle\psi|\hat A|\psi\rangle$, where $\hat A$ is a unitary operator acting on $\ket{\psi}$ and the ancilla qubit is initially prepared in $\ket{0}$. At the end of the circuit, the ancillary qubit is measured to obtain $\langle\hat{\sigma}^z-i\hat{\sigma}^y\rangle$. The expectation of $\hat A$ follows from $\bra{\psi}\hat A\ket{\psi} =  \langle\hat{\sigma}^z-i\hat{\sigma}^y\rangle$. (d) An alternative path to measure the $S$-matrix. The path can be used to measure an unknown anyon (blue) at a fixed position by braiding with anyons of known species (black). }
    \label{braiding_general}
\end{figure}

\section{Anyonic Braiding and Fusion Channels}\label{anyonbraiding}
Anyons emerge as the non-local excitations of topologically ordered phase. They are quasiparticles  that obey fractionalized statistics. The existence of these exotic anyonic excitations is a defining property of topologically ordered phases. Two key quantities of the underlying anyonic theories are the twist factors $\theta_a$ and modular $S$-matrix. They can be used to classify the theory down to finitely many possibilities \cite{etingof:2005}  (up to gauge equivalence). The measurement of these data serves as another useful tool to characterize the topological order experimentally~\cite{Parsa:2021}.

The anyonic twist factor $\theta_a$ is defined to be the phase accumulated when a particle is rotated by $2\pi$ about itself. For abelian anyons (including bosons and fermions), the twist factor determines their braiding statistics, i.e. the phase resulting from exchanging two identical particles. Whereas $\theta_a=\pm 1$ for bosons and fermions, it can take other rational phases for generic anyons. To obtain the twist factor, we first create a pair of anyons, $a$ and  $\Bar{a}$ (particle and anti-particle), from the ground state and then move particle $a$ to the final position with unitary $\hat{U}_a$ along a twisted path as shown in Fig.~\ref{braiding_general}b. The twist factor is the phase relative to the state with the particle moved to the same location along a path without a twist. We denote the unitary that drags particle $a$ along this alternative (untwisted) path as $\hat{V}_a$.

The modular $S$-matrix captures the mutual statistics between the anyons. In the abelian case, the matrix elements encode the phase accumulated when one anyon winds around another. In the non-abelian case, the winding induces a non-trivial transformation in the subspace formed by these anyonic particles.  To measure the $S$-matrix, we first create two pairs of anyons $a,\Bar{a}, b,\Bar{b}$ from the ground state and then intertwine $a$ and $b$ with unitary $\hat{U}_{ab}$ (see Fig.~\ref{braiding_general}a). Next we consider another path corresponding to moving these anyons to the same location without crossing each other along the way; we denote the unitary for this path as $\hat{V}_{ab}$.
Matrix element $S_{ab}$ is obtained by measuring the overlap between the final states in the two scenarios above. The twist factor $\theta_a$ and the $S$-matrix are expressed in terms of expectation of unitaries $\hat{V}_{a}$, $\hat{U}_{a}$ and $\hat{U}_{ab}$, $\hat{V}_{ab}$ as follows:
\begin{align}
    \theta_a &= \bra{\psi_{a\Bar{a}}}\hat{V}_{a}^{\dag}\hat{U}_{a}\ket{\psi_{a\Bar{a}}}, \nonumber \\
    M_{ab} &=
    \bra{\psi_{a\Bar{a}b\Bar{b}}}\hat{V}^{\dag}_{ab}\hat{U}_{ab}\ket{\psi_{a\Bar{a}b\Bar{b}}} \label{eq:braiding}.
\end{align}
Here $M_{ab}$ is the monodromy matrix \cite{bonderson:2006} and it is related to the $S$-matrix via $S_{ab} =  \left({d_ad_b}/{\mathcal{D}}\right)M_{ab}$, where $d_{a,b}$ is the quantum dimension of anyons $a,b$. $\ket{\psi_{a\Bar{a}}}$ is the wavefunction with a pair of anyons $a,\Bar{a}$ and $\ket{\psi_{a\Bar{a}b\Bar{b}}}$ is the wavefunction with two anyon pairs $a,\Bar{a},b,\Bar{b}$. Once we know how to implement $\hat{U}$ and $\hat{V}$, the expectation value can be efficiently measured by a simple Hadamard-test quantum circuit with one ancilla qubit as shown in Fig.~\ref{braiding_general}c. The number of local gates needed to perform the measurement will scale linearly with the number of qudits supported by the unitary being measured. The costs come from the long-range controlled gates from the ancilla qubit to the support of the unitary. The procedure is a typical example of Mach-Zehnder or Ramsey type interferometry for measuring the braiding statistics \cite{bonderson:2006, Jiang:2008} where the anyonic statistics results from the interference of paths $\hat{V}$ and $\hat{U}$. 

Notice that the details of the paths in Fig.~\ref{braiding_general}a-b are not important, any paths that can be continuously deformed to them are considered equivalent. 
Some non-universal phases can emerge locally along the paths (e.g. geometric phases). Nevertheless, the braiding phase can still be determined by splitting a single path into multiple segments. These segments of paths can be implemented in different order to create trajectories with and without the particle braiding, the interference of which determines the statistics~\cite{levin:2003}. In our approach we do not encounter these non-universal phases since we can explicitly construct the Wilson string operators corresponding to the anyons, as explained in Section~\ref{sec:string_op}.

A measurement of the monodromy matrix alone is sufficient for determining the modular $S$-matrix for string-net models.
This is because the modular $S$-matrix is required to satisfy certain constraints, e.g. the modular $S$-matrix is symmetric and unitary, and satisfies the Verlinde formula \cite{verlinde:1988}. These stringent constraints do not only allow us to directly infer the modular $S$-matrix from the monodromy matrix measured from the experiments,
\begin{equation}
     S_{ab} = \sqrt{\frac{(M^{-1})_{ba}^*}{M_{ab}}}M_{ab},
\end{equation}
where $M^{-1}$ is the inverse of $M$.
In practice they are helpful in identifying the correct modular $S$-matrix from the noisy data (see Appendix~\ref{append:smatrixfrommea}).

Measurement of these interfered paths also provides access to the {\it fusion channel}. If we bring two abelian anyons together, they can be regarded as a single particle of a unique type. Abelian anyons have a unique fusion channel whereas non-abelian anyons have multiple fusion channels by definition. For non-abelian anyons, combining them can result in a particle of multiple types. Suppose that we have an unknown anyonic state $\ket{\psi} = \sum_i p_i\ket{i}$ at a fixed position, where $i$ labels all the possible anyons and $p_i$ are some complex amplitudes. We can prepare a known anyon $a$ and braid it around the unknown anyon (Fig.~\ref{braiding_general}d); interfering this winding path with the path without winding gives the same quantity as the $S$-matrix element in \eqref{eq:braiding}, i.e. $\bra{\psi_a}\hat{V}^{\dag}\hat{U}\ket{\psi_a} = \mathcal{D}\sum_i |p_i|^2S_{ai}/d_ad_i$, where $\ket{\psi_a}$ is $\ket{\psi}$ with a pair of $a,\bar{a}$ created and $\hat U,\hat V$ are the unitaries for the two paths. The probability amplitudes $|p_i|^2$ are determined by inverting the matrix $S_{ai}/d_a d_i$. Measurement of the fusion results allows us to directly access the fundamental algebraic relation underlying the TQFT~\cite{nayak:2008}. 

\begin{figure}
    \centering
    \includegraphics{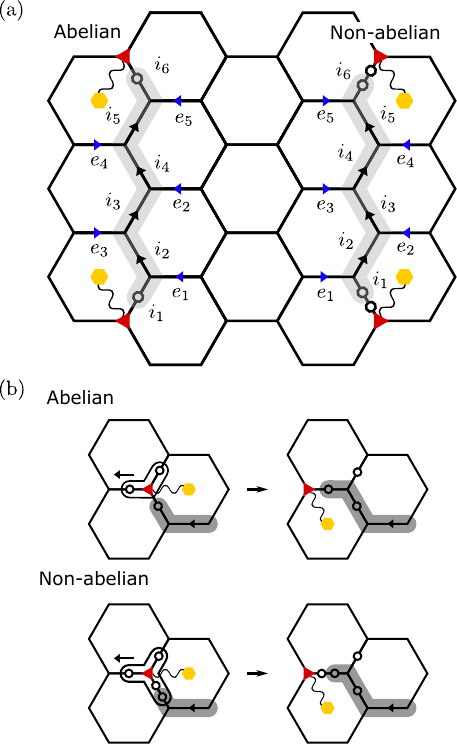}
    \caption{Prepare and move anyons with unitary string operators. (a) Open string operators for abelian (left) and non-abelian (right) anyons. The operators act on the set of labeled qudits $\{i\}$ along the path and $\{e\}$ of the legs. Only the states of the qudits $\{i\}$ along the path are changed by the operator. Each quasiparticle has an associated vertex (red) and plaquette (yellow). Before the string operators are applied, the ancillary qudits in the non-abelian case are initialized to align with the other qudits on the same edge. (b) The endpoint quasiparticles can be moved by extending the string operator sequentially with unitary gates. The dark gray marks the existing string, The enclosed qudits on the left are acted on by the endpoint-moving unitary gates, which correspond to an exact (modified) form of the unitary string operators for abelian (non-abelian) case. The plaquette-vertex labels are used to locate the quasiparticle. When extending the string operator, they trace the motion of the quasiparticle. Note that the ancillary qudit in the non-abelian case can be moved without physically displacing the qudits. See Appendices~\ref{append:stringop} and ~\ref{append:explicitcircuit} for more details.}
    \label{string_op}
\end{figure}

\begin{figure}
    \centering
    \includegraphics{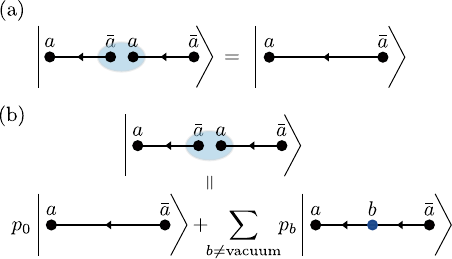}
    \caption{The manipulation of abelian and non-abelian anyons. (a) Two separate abelian anyons $a,\bar{a}$ can be created by a sequence of multiple local anyon-antianyon pairs. The intermediate quasiparticles fuse to the vacuum. (b) If $a,\bar{a}$ are non-abelian anyons, the transport of the anyon is no longer equivalent to local pair creation and pair annihilation of $a$ and $\bar{a}$. The fusion yields a superposition of the vacuum and some other non-trivial anyon species $b$, with amplitudes $p_b$.}
    \label{fig:ab_nonab_diff}
\end{figure}
\section{Creating and Moving Anyons in the String-net Model}\label{sec:string_op}
In this section, we describe how abelian anyons can be created and braided using quantum circuits of a constant depth, whereas braiding of non-abelian anyons requires linear-depth (in anyon--anyon separation) quantum circuits due to their non-unique fusion.

The quasiparticles in the string-net model are identified with closed string (Wilson loop) operators that commute with the string-net Hamiltonian. When a closed string operator is broken to have open ends, the quasiparticle and the corresponding anti-quasiparticle emerge as defects localized at the endpoints of the open string operator applied to the string-net states~\cite{levin:2005}. The string operators connecting the two endpoints can be thought of as a trajectory traced by the quasiparticle at one end while the other is at rest. As a result, the braiding events can be simulated by realizing the open string operators with a sequence of unitary gates. 

The precise details of how the closed string operators are broken into open strings are not essential for our purpose. Braiding is described by the relative motion between different quasiparticles. In other words, the braidings are captured by how the open string operators cross each other away from their endpoints (see Fig.~\ref{braiding_general}). We require the open string operators to have the exact same bulk form as the closed string operators given in Ref.~\cite{levin:2005}, but at the same time to be unitary. These string operators are generally isometries, but can be promoted to a unitary form. Such \emph{unitary string operators} allow the manipulation of anyons by decomposition into a sequence of smaller gates.

To simplify the discussion, we focus on the string-net models where quasiparticles correspond to either a simple string operator or a product of simple string operators. This is not a very restrictive condition and, for instance, includes all of the spin-1/2 string-net models originally considered in Ref.~\cite{levin:2005} and non-abelian anyons beyond spin-1/2 lattice such as the Ising anyon.
A string operator being {\it simple} means that it can be assigned a unique basis string label $s\in\{0,1,...,N\}$ and it characterizes an irreducible quasiparticle in the model
(see Appendix~\ref{append:stringnetsummary} for details). We give a definition of the open simple string operators that can be promoted to a unitary in Appendix~\ref{append:stringop}. The shapes of the operators are depicted in Fig.~\ref{string_op}a for general abelian and non-abelian cases. The string operator changes the qudit states $\{i\}$ along its path and leaves the states $\{e\}$ on the legs unchanged. The orientation of the path along the string is indicated with arrows. The quasiparticles at the endpoints of the open string can be moved by deforming the string operator with some local unitary gates (see Fig.~\ref{string_op}b). The product of simple strings can be implemented by multiple simple strings with non-overlapping endpoints.

\subsection{Locating the quasiparticles}
\begin{figure}
    \centering
    \includegraphics{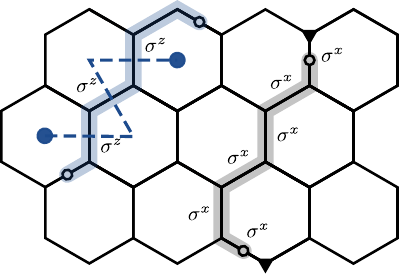}
    \caption{The string operators for toric code. The end points can be labeled by a plaquette or vertex that corresponds to exactly the microscopic excitation in Eq.~\eqref{HTC}. The dashed line indicates the path that the plaquette excitation takes as the Wilson string is generated.}
    \label{TC_example}
\end{figure}

The braiding interferometry in Section~\ref{anyonbraiding} assumes the quasiparticle configurations at the end of the interfered paths to be identical. To unambiguously locate a quasiparticle at the endpoint of a string operator, we label each endpoint with a vertex and a plaquette. If the string operator turns left (right) before the spin at the endpoint, then the associated plaquette is on the left (right) of the endpoint. The associated vertex is the vertex away from the endpoint of the string operator. The plaquette-vertex pair labels the location and orientation of the quasiparticle (see Fig.~\ref{string_op}a). The motion of the quasiparticle can be visualized as moving the pair of labels as shown in Fig.~\ref{string_op}b. Note that the plaquette/vertex label does not necessarily coincide with the actual plaquette-vertex violation that appears as excitation in the microscopic string-net Hamiltonian. In some cases, they can be chosen to coincide. An example is the TC excitations which are created by string of $\hat\sigma^x$ along the vertex labels and $\hat\sigma^z$ along the plaquette labels (Fig.~\ref{TC_example}). Each string operator thus only needs one of the plaquette or vertex labels and the other one becomes irrelevant.

\subsection{Unitary string operators for the abelian anyons}
If the quasiparticle describes an abelian anyon, the trajectory represented by the string operator can be interpreted as a chain of anyon--antianyon pair creations. Since the abelian anyon and its antianyon are guaranteed to fuse into the vacuum, the result will be the same as having one quasiparticle at each end of this chain, as depicted in Fig.~\ref{fig:ab_nonab_diff}a.

\begin{figure}
    \centering
    \includegraphics{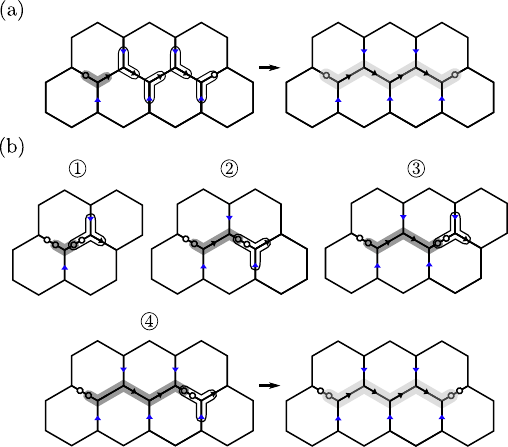}
    \caption{The decomposed string operators using the movements from Fig.~\ref{string_op}b. (a) For the abelian string operators, an initial short string (dark gray) can be extended to a long string operator (light gray) by a constant-depth local quantum circuit. (b) A sequentially applied quantum circuit is required for extending a short non-abelian string, resulting in a linear-depth quantum circuit. We show the details of the circuits, including the short string initialization in Appendix~\ref{append:stringop}.}
    \label{fig:stringop_decomp}
\end{figure}

This shows that a long unitary string operator for the abelian anyons is a product of disjoint short unitary string operators. 
As seen in the circuit decomposition of the string operator as shown in Fig.~\ref{string_op}b, an abelian open string operator can be extended by one more site,  with a two-qudit quantum gate that does not overlap with the existing string operator. In this case, this two-qudit gate corresponds to a unitary open string operator in Fig.~\ref{string_op}a with one endpoint slightly modified.
To create a long string operator of the form in Fig.~\ref{string_op}a, we can apply a short string operator (a three-qudit quantum gate) and extend it to a longer string as schematically illustrated in Fig~\ref{fig:stringop_decomp}a. Since none of the gates overlap with each other, the circuit can be implemented with a constant depth. Further details of the two- and three-qudit gates are given in Appendix~\ref{append:stringop}.

\subsection{The difficulty of manipulating non-abelian anyons}
The argument given for the abelian anyons no longer holds when we consider the non-abelian case. The fusion of two non-abelian anyons along the chain does not give only the vacuum, but in general a superposition of fusion channels (see Section~\ref{anyonbraiding} and Fig.~\ref{fig:ab_nonab_diff}b). To create two separated non-abelian anyons from the local pairs, additional projections are needed around the anyon-antianyon pair to ensure that they fuse to the vacuum.

Therefore, a unitary (and hence reversible) movement of non-abelian anyons can only be achieved sequentially, requiring at least a number of time steps that scales with the separation between the anyons. In the language of field theories, this amounts to the necessity of path-ordering in the Wilson loop operator for non-abelian anyons. This intuition implies that at least a linear-depth quantum circuit is needed to implement a unitary string operator that describes the movement of non-abelian anyons. A recent work~\cite{bravyi:2022} has proved this lower bound in the context of the quantum double model.

\subsection{Unitary string operators for non-abelian anyons}
To define a unitary string operator for the non-abelian anyons, we introduce an additional qudit at each endpoint of the string as shown in Fig.~\ref{string_op}a. This additional degree of freedom at the string endpoints is motivated by the diagrammatic calculus of the string operators in the Appendix D of Ref.~\cite{levin:2005}. In the diagrammatic approach, an endpoint of a string operator will carry a simple string label and split the edge it touches (represented by a physical qudit) into two edges; One of the edges still retains the original qudit state, and the other edge now stores the new qudit state obtained from applying the string operator. In our protocol, we introduce the additional qudit to take into account this additional split edge so that we can keep the full information about the open string endpoint, making the unitary construction of the operator possible.

As depicted in Fig.~\ref{string_op}b, a non-abelian open string operator can be extended by one site with a four-qudit quantum gate. In contrast with the abelian case, the gate that extends the string operator needs to involve the qudits at the endpoint of the existing string operator, and the gate itself cannot be regarded as a short unitary open string operator. To create a long string operator, we first implement a short unitary string operator of the form in Fig.~\ref{string_op}a (in the non-abelian case, this is achieved by a four-qudit quantum gate as explained in Appendix~\ref{append:stringop}), the short string is extended using a sequence of four-qudit gates in Fig.~\ref{string_op}b. An explicit procedure is depicted in Fig.~\ref{fig:stringop_decomp}b, which is a quantum circuit the depth of which grows linearly with the separation of the string endpoints.
Note that the resulting unitary string operators satisfy the optimal scaling of the circuit depth argued above. The existence of the efficient string operator decomposition allows for manipulation of non-abelian anyons without the need of any projection.

When simulating the non-abelian strings, the additional ancillary qudits should always be placed at the endpoints of the current string operators. This can be achieved by swapping the ancilla states through the lattice without physically displacing the qudits. We provide full details of the protocol together with explicit quantum circuits in Appendices~\ref{append:stringop} and~\ref{append:explicitcircuit}. Alternatively, we can simply place two qudits on each edge instead of only one. Despite doubling the number of qudits, the alternative approach retains the translational and rotational invariance of the lattice.


\begin{figure}
    \centering
    \includegraphics{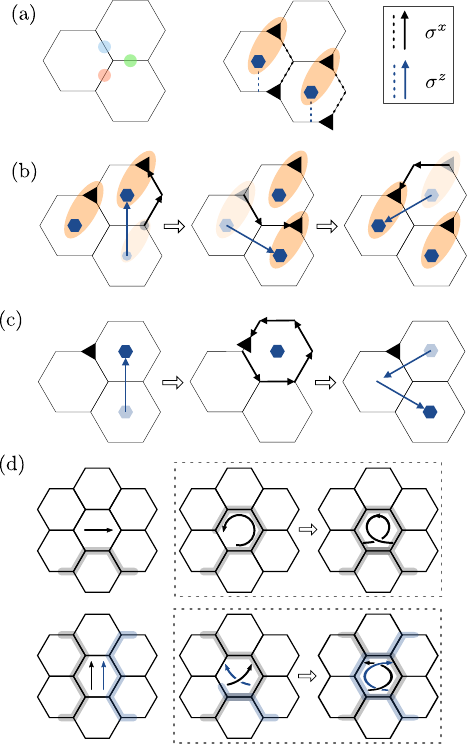}
    \caption{Small-scale experimental realization. (a) In a three-plaquette toric code, TEE can be measured using the three connected domains as in Fig.~\ref{tee}, consisting of 3 qubits. On the right we show the initial configuration of the anyons prepared by $\hat{\sigma}^x$ and $\hat{\sigma}^z$, the anyons can also be moved using the same operators. (b)
    Braiding steps to realize the exchange of $e\times m$ particles. The solid lines are string operators used to drag the particles. (d) We can extract other $S$-matrix elements by following (c) twice with some specific anyons removed. Here we only keep the $m$ on the left and $e$ on the right. Applying (c) twice results in three parallelized steps as shown, which gives $S_{em} = S_{me} = -1$. (d) The braiding for general abelian and non-abelian anyons can be simulated on a small system as shown. The left and right diagrams depict the interfered paths of the anyons. Note that the final anyon configurations are identical.}
    \label{minimal_TC}
\end{figure}
\section{Minimal Experimental Realization on NISQ device}\label{section:exp}
In the era of NISQ computation, the major challenge comes from the intrinsic noise and the various imperfections in state initialization and gate operations. A feasible experimental realization of a topologically ordered state must be conducted with shallow circuits of local gates and efficient braiding that can be used to measure all the twist factors and the full $S$-matrix. The scheme described above yields promising designs for small-scale experiments that are already possible on NISQ devices, in particular for the abelian spin-1/2 models (TC and DS). 

A minimal example for the experiment of a three-plaquette TC is depicted in Fig.~\ref{minimal_TC}a (see also~\cite{Satzinger2021} for a square lattice version). The TEE can be inferred from minimum system of three qubits around a vertex using the subtraction procedure in Section~\ref{sec:tee}. The quasiparticles in TC come in four types, the trivial $\mathbf{I}$, $e$ (string of $\hat \sigma^x$), $m$ (string of $\hat \sigma^z$) and the composite $e\times m$ (product of $e$ and $m$). They can be prepared on the TC ground state using strings of Pauli matrices. To understand the braiding in TC, it suffices to consider only the braiding paths for the composite $e\times m$, from which all the other paths of the braiding can be extracted. Fig.~\ref{minimal_TC}b shows the path to exchange two $e\times m$ particles, where we make use of the property that the twist factor for abelian anyons can be extracted from the phase when exchanging two identical particles. The mutual statistics can be obtained by interchanging the spatial positions of the particles twice and back to the initial quasiparticle configuration. It is not hard to check that the paths traced out by the particles are the same as Fig.~\ref{braiding_general}a up to a smooth deformation. An example of the mutual braiding between $e$ and $m$ is shown in Fig~\ref{minimal_TC}c.

For more general abelian and non-abelian models, the braiding can be carried out on a minimal systems such as that shown in Fig.~\ref{minimal_TC}d. The system is subject to further reduction on the number of the boundary qudits that can be disentangled without affecting any observables in the bulk. Using the protocols that we have, it would be exciting to realize a proof-of-principle simulation of the non-abelian spin-1/2 double Fibonacci model (DF) that supports the non-abelian Fibonacci anyon as quasiparticle excitation. Apart from its exotic non-abelian physical nature, the Fibonacci anyons are also known to be universal in topological quantum computation~\cite{nayak:2008}. In Appendix~\ref{append:dfibexample}, we provide the circuit diagrams for the state preparation and anyonic braiding in DF. The circuits yield a rough estimate for the resources overhead of DF compared to the TC and DS based on the C-$\hat B_p$ operator in the state preparation, which can be decomposed into $\gtrsim 5$ local CNOTs for TC and $\gtrsim 15$ local CNOTs for DS, but $\gtrsim 120$ CNOTs for DF. In other words, to construct a string-net ground state on a system of $N\times M$ hexagon plaquetes ($\sim 3NM$ qubits) and $N\leq M$, the layers of parallel local CNOT gates scale as $\gtrsim 5N$ for TC, $\gtrsim 15N$ for DS and  $\gtrsim 120N$ for DF. Although the resources needed to realize the DF are an order of magnitude more demanding than the abelian examples, the quests for small-scale experiments have already begun~\cite{li:2017}.

\section{Discussion}
In light of the ground state structure of the string-net models, we develop a set of efficient quantum circuits for preparing the topological string-net states with depth that scales linearly as the minimum width of the system. The implementation offers substantial practical advantages compared to the previously known unitary constructions~\cite{konig:2009,Soejima2020}.
The prepared topological states serve as platforms supporting abelian or non-abelian anyons, which are braided by the unitary string operators. 
Although we focus our discussion on the honeycomb lattice, the implementation of the protocol can easily be run on the heavy-hexagon lattice~\cite{chamberland:2020}, which is available in current quantum computing platforms. The construction also straightforwardly generalizes to any other 2D trivalent lattices compatible with the architecture of the devices, rendering further reduction on the number of qudits needed. The explicit quantum gates that we use during the construction are local within each plaquette on the lattice (e.g. Fig.~\ref{general_sn}). In practice, the implementation of these gates needs to be optimized based on specific device connectivity, as has been done in Ref.~\cite{Satzinger2021} for the square lattice.

The circuit construction can be extended to other topological states or quantum stabilizer codes that share a similar ground state structure. An example is Kitaev's quantum double model~\cite{kitaev:2003}, which can be defined on any lattice embedded in 2D orientable surface with orthonormal basis $\{\ket{g}|g\in G\}$ labeled by the elements of a finite group $G$. The Hamiltonian also takes the form as a sum of commuting local projectors on vertices and plaquettes. We can use similar idea to obtain a linear quantum circuit for its ground state with the representative spin in each plaquette initialized as $\frac{1}{|G|}\sum_{g\in G}\ket{g}$. This is not surprising as it is known that quantum double models can be mapped to a subclass of string-net models \cite{buerschaper:2009,kadar:2010}. 
The same scheme can also be used to prepare any states that are related to the string-net states by a finite-depth quantum circuit, the corresponding quasiparticle string operators being smeared out with support bounded by the light cone of the finite circuit. The states of this form (often with a finite correlation length) will exhibit the same topological order~\cite{gu:2010}, making them valuable computing resources for the study of correlated quantum many-body physics with controlled quantum systems.

The preparation scheme can be exploited as an efficient generator for a family of 2D quantum datasets on the digital quantum computers. It would be interesting to use these exotic states for benchmarking various protocols where the non-trivial order of the states are relevant. Some examples include quantum phase recognition~\cite{cong:2019,smith:2020}, quantum tomography for weakly entangled states~\cite{cramer:2010, lanyon:2017, deng:2017} or entanglement measurements~\cite{vanEnk:2012,Elben:2018,brydges:2019,Huang:2020}, which have so far mostly only been illustrated  on 1D cases. These quantum data can also be efficiently generated on a photonic quantum computer~\cite{wei:2021a}.

Another exciting direction to be explored is fault-tolerant quantum computation. The circuits realize an efficient unitary encoding of the logical information in planar geometry with suitable boundary conditions, i.e. the surface code protocol~\cite{fowler:2012, eric:2002,horsman:2012}. Despite not being intrinsically fault-tolerant, small-scale unitary encoding already provides useful insights into the logical state injection and logical error dynamics~\cite{Satzinger2021}. An alternative approach to achieve fault-tolerance is the topological quantum computation (TQC)~\cite{nayak:2008}, where logical information is encoded in non-abelian anyons. 
A key advantage of TQC compared to the surface code protocol is that it allows for \emph{universal} fault-tolerant computation by anyonic braiding. While universal computation is possible in surface code, it requires a costly state distillation process. The unitary quasiparticle string provides a unitary encoding scheme for TQC (e.g. based on Fibonacci anyons). The braiding is accomplished by a sequence of local gates, which can also be used to determine the fusion of the anyons. 
However, whether all these anyonic operations can be performed fault-tolerantly, e.g., by introducing syndrome measurements~\cite{bonesteel:2012, Schotte2020} or anyon distillation~\cite{konig:2010}, remains an interesting open question for future work.

\section{Acknowledgement}
We thank Parsa Bonderson, Robert K{\"o}nig and Zhiyuan Wei for helpful discussions. Y.-J.L was supported by the Max Planck Gesellschaft (MPG) through the International Max Planck Research School for Quantum Science and Technology (IMPRS-QST). A.S. was supported by a Research Fellowship from the Royal Commission for the Exhibition of 1851. F.P. acknowledges support from the European Research Council (ERC) under the European Union’s Horizon 2020 research and innovation programme (grant agreement No. 771537) and the Deutsche Forschungsgemeinschaft (DFG, German Research Foundation) under Germany’s Excellence Strategy EXC-2111-390814868 and TRR 80. K.S. was supported in part by the US-Israel Binational Science Foundation under Grant No.~2016255. 
\bibliographystyle{apsrev4-2}
\bibliography{references.bib}

\appendix
\section{A Brief Summary of String-net Model}\label{append:stringnetsummary}
Here we give a brief summary of the string-net model introduced in \cite{levin:2005}. There exist generalizations beyond the original construction \cite{kong:2013,lan:2014,LinLevin:2014}. In this paper we focus on the original construction in (2+1)D with rotational invariance. 

\begin{figure}
    \centering
    \includegraphics{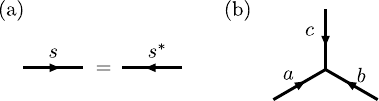}
    \caption{(a) Inverting a string $s$ is the same as changing to the dual string $s^*$. (b) The orientation convention for branching at a vertex.}
    \label{fig:orientation}
\end{figure}

The string-net models are defined on a 2D trivalent graph where each edge is associated with a string. We label the strings by $s = 0,1,2,...,N$ where $N$ is the total number of nontrivial string types in the model, with $s = 0$ labeling the absence of any strings. The strings are generally oriented, each string $s$ has a dual string $s^*$ which points to the opposite direction (see Fig.~\ref{fig:orientation}a). As a result, the unoriented string satisfies $s = s^*$. The branching rule is the set of triplets of strings $\{a,b,c\}$ allowed to meet at the vertex, with the orientation convention shown in Fig.~\ref{fig:orientation}b. We define the branching delta $\delta_{abs} = 1$ if the branching $\{a,b,c\}$ is allowed, and zero otherwise. The null string $s = 0$ is associated with the branching rule $\delta_{0ss^*} = 1$ for any string type $s$. In the abelian string-net model (the excitations are abelian anyons), if two strings $a,b$ meet (branch) at a vertex, then the third string $c$ at the same vertex is uniquely determined \cite{LinLevin:2014}, we denote this by $c^* = a\times b$. For a continuous string $\{0,s,s^*\}$ we have $s = s\times 0$. For general non-abelian string-net models, the third string $c$ is not uniquely determined.

The string-net ground state is the state that satisfies a set of local rules. More precisely, let us consider a configuration $X$ in the ground state wavefunction, its amplitude is denoted by $\Phi(X)$. The ground state is the weighted superpositions of string configurations $\ket{\psi} = \sum_X\Phi(X)\ket{X}$, The set of local rules relate the weights of the configurations that only differ locally
\begin{equation}\label{localrules}
    \includegraphics[scale = 1]{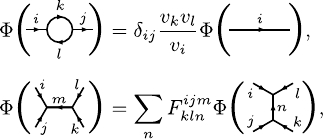}
\end{equation}
where $v_s = \sqrt{\pm d_s}$ with $d_s>0$ being the quantum dimension of string $s$, they also satisfy $v_0 = 1$ and $v_s = v_{s^*}$. The rank-six tensor $F$ is called the $F$-symbol. For convenience we usually define $F^{ijm}_{kln}$ to be zero if the corresponding diagram has forbidden branching. By setting $i=j = 0$, we see that a type-$s$ loop carries a weight of $b_s = v_s^2\in\mathbb{R}$. These rules completely determine the ground state wavefunction in the sense that the amplitude of any configuration can be reduced to the amplitude of trivial configuration $\ket{00...0}$ by applying these rules (neglecting the boundary conditions).

Because one configuration can be related to the same configuration by applying the local rules in different order, they therefore satisfy the so-called pentagon equation:
\begin{equation}\label{pentagon}
    \sum_{n} F^{mlq}_{kp^*n}F^{jip}_{mns^*}F^{js^*n}_{lkr^*} = F^{jip}_{q^*kr^*}F^{riq^*}_{mls^*}.
\end{equation}
In addition, it can be shown that in order for the string-net model to be physical and self-consistent, $F$-symbol should satisfy
\begin{align}\label{consistency}
     F^{i^*j^*m^*}_{k^*l^*n^*} &= \left(F^{ijk}_{kln}\right)^*, \nonumber\\
     F^{ijm}_{j^*i^*0} &= \frac{v_m}{v_iv_j}\delta_{ijm}, \nonumber\\
     F^{ijm}_{kln} = F^{lkm^*}_{jin} &= F^{jim}_{lkn^*} = F^{imj}_{k^*nl}\frac{v_mv_n}{v_jv_l},
\end{align}
where the star $*$ on the scalar denotes the complex conjugate of that number. The algebraic object $(F^{ijk}_{kln},b_s)$ that satisfies the constraints above is what defines a string-net model. Using the equations from above, we can derive one of the most useful properties of the $F$-symbol that underlines the unitarity of our construction in the main texts
\begin{equation}\label{unitarity}
    \sum_{n} F^{ijm}_{kln}\left(F^{ijm'}_{kln}\right)^* =\delta_{mm'} \delta_{ijm}\delta_{klm^*}.
\end{equation}
Another useful identity that can be drived from Eqs.~\eqref{pentagon} and \eqref{consistency} is
\begin{equation}\label{append:eq:fusion}
    \sum_s \delta_{abc^*}d_c = d_ad_b,
\end{equation}
where $d_s = |v_s|^2>0 $ turns out to be the quantum dimension of the anyon realized by the type-$s$ simple string operator.

The sring-net wavefunction is the ground state of the Hamiltonian \eqref{H}, where the plaquette operator is written as $\hat{B}_p = \sum_s a_s\hat{B}^s_p$. For the topological phase to have a continuum limit, $a_s$ is chosen to be $a_s = b_s/\sum_sb_s^2$. With this choice of $a_s$, $\hat{B}_p$ becomes a projector. Each $\hat{B}^s_p$ physically corresponds to adding a type-$s$ loop to a plaquette and fusing it into the lattice based on the local rules Eq.~\eqref{localrules}. With this graphical picture, one can work out the general matrix element for $\hat{B}^s_p$ in terms of the $F$-symbol
\begin{equation}\label{b_define}
    \includegraphics[scale = 1]{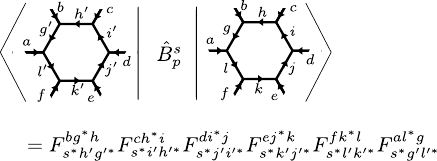}.
\end{equation}
Note that the spins on the six external legs are not changed upon the application of $\hat B_p^s$. The operators of any string types and on any plaquettes commute, i.e. $[\hat{B}^{s_1}_{p_1},\hat{B}^{s_2}_{p_2}]=0$ for any $s_1,s_2$ and $p_1,p_2$.

Each quasiparticle excitation of the Hamiltonian \eqref{H} corresponds to a closed string operator. The string should not be observable, only the endpoints of the string (quasiparticles) are. For this reason, the closed string operator should commute with the Hamiltonian. A type-$s$ string operator that fulfills the requirements has the matrix element from $i$ to $i'$
\begin{equation}\label{append:eq:string_op}
     W^{i'_1i'_2...}_{i_1i_2...}(e) = \left(\prod_{k }F^{e_{k-1}i_k^*i_{k-1}}_{s^*i'_{k-1}i'^*_k}\right)\left(\prod_{k } \omega_k\right),
\end{equation}
where $i_1,i_2,...$ are the spins along the closed string, $e$ is the set of external legs along the path of the string operator (e.g. see Fig.~\ref{string_op}a). The closed string operator $W$ only changes the spins along the path but not the ones on external legs. The quantity $\omega$ is defined as
\begin{equation}
    \omega_k = 
    \begin{cases}
    \frac{v_{i_k}v_s}{v_{i'_k}}\omega^{i'_k}_{i_k}, &\text{if before/after }i_k\text{, it turns right/left,} \\
    \frac{v_{i_k}v_s}{v_{i'_k}}\overbar{\omega}^{i'_k}_{i_k}, &\text{if before/after }i_k\text{, it turns left/right,} \\
    1, &\text{otherwise}.
    \end{cases} \label{omega}
\end{equation}
where $\omega^i_j$ and $\overbar{\omega}^i_j$ are phase factors satisfying Eq.~\eqref{appendix:eq:omega}. Notice that when $W$ acts on a plaquette $p$, it should correspond to adding a type-$s$ string loop and fusing it into the plaquette, i.e. it is equivalent to applying $\hat{B}^s_p$.

The solutions of $\omega$ are solved from a set of consistency equations by imposing the commutativity of the string operators with the Hamiltonian  
\begin{align*}
\overbar{\omega}^j_i &= \sum_n\omega^k_{i^*}F^{is^*k}_{i^*sj^*},
\\
\frac{v_sv_j}{v_m}\overbar{\omega}^m_jF^{sl^*i}_{kjm^*}\omega^l_j &= \sum_n F^{ji^*k}_{s^*nl^*}\omega^n_kF^{jl^*n}_{ksm^*}.
\label{appendix:eq:omega}
\numberthis
\end{align*}
The string operator of this form is called the \emph{simple string operator}. The product of two simple string operators $W_1W_2$ results in another string operator for the composite quaisparticle. The simple string operators form a large subclass of the general string operators, they describe many interesting anyonic quaisparticles that are relevant for physical realizations, such as the anyons in the abelian TC and DS, and the non-abelian double Fibonacci model.
We speculate that it is also possible to generalize the unitary definition of the open string operators to the non-simple cases using ancillary qudits, this is discussed in Appendix~\ref{append:isometry}.

\section{Quasiparticle string opeartors}\label{append:stringop}
The anyonic excitations in the string-net models are localized at the endpoints of the open string operators. There is no unique way to define the endpoints (excitations) of these open string operators. However, the topological properties of these quasiparticles are independent of such details. Here we give a definition of the open string operator (corresponding to a type-$s$ string of length $L$) in the string-net model, its matrix element from initial spins $i_1,i_2,...$ to $i_1',i_2',...$ can be written as
\begin{equation}
    W^{i'_1i'_2...i'_L}_{i_1i_2...i_L}(e) = \frac{v_{i'_1}}{v_{i_1}v_s}\left(\prod^{L}_{k = 2}F^{e_{k-1}i_k^*i_{k-1}}_{s^*i'_{k-1}i'^*_k}\right)\left(\prod_{k = 2}^{L-1} \omega_k\right) \label{os},
\end{equation}
where $e = \{e_1,e_2,...e_{L-1}\}$ is the set of external legs along the string (see Fig.~\ref{string_op}a). The complex numbers $v_i$ and rank-six tensor $F$ are the data that define the given string-net model (see Appendix~\ref{append:stringnetsummary}). $\omega$ is defined in Eq.~\eqref{omega}.

This open string operator has the property that creating a type-$s$ string along $\{i_1,i_2,...,i_L\}$ is the same as creating a type-$s^*$ string along $\{i_L,...,i_2,i_1\}$. In addition, the open string operator satisfies an isometry condition that underlies the unitary decomposition of the operator as discussed below.

{\bf Abelian quasiparticle strings:} \label{appendix:string_op}
In the abelian theory, we can move the abelian anyon by applying the open string operator that connects between the anyons at the initial and final position. That is, we change the position of the endpoint by joining two open strings together to form a new open string. This is possible because abelian anyons have a unique fusion channel. When two endpoints (anyons) join, the anyon and its anti-partner combine (fuse) to vacuum. 

Since we can move the anyon using a sequence of local unitary, it is convenient to initialize a short open string operator and move the anyons (endpoints) apart from each other later. We can prepare a pair of abelian anyons (that corresponds to an type-$s$ string) by a three-qudit unitary having the matrix elements in~\eqref{os} for $L = 2$
\begin{equation}
    \includegraphics[scale = 1]{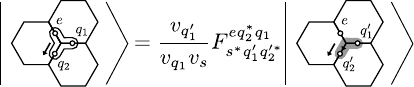},
    \label{prepabelian}
\end{equation}
where $q'_{1,2} = s\times q_{1,2}$ and $\{e,q_1,q_2^*\}$ is an allowed branching. The dark path on the right hand side indicates the path created by the string operator. We use arrow along the string to indicate the orientation convention following Fig.~\ref{string_op}a. In abelian theory, the non-trivial values in $F$-symbol become phase factors (see Appendix~\ref{append:stringnetsummary}). This operator initializes a string of length two on the lattice. Note that we could have set the phase $v_{q'_1}/v_{q_1}v_s$ to 1 by local unitary in this abelian case, but in non-abelian theory this factor is no longer a phase and it becomes important for preserving the isometry property of the open string operator.

\begin{figure}[t]
    \centering
    \includegraphics{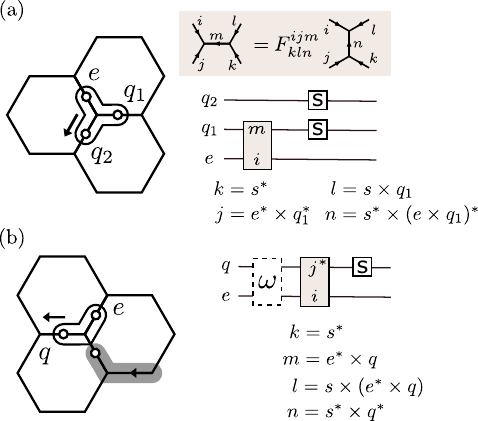}
    \caption{The quantum circuits for abelian open string operators. (a) A circuit realization for preparing abelian anyons. The shaded box shows the $F$-move defined by the $F$-symbol in the abelian model, together with the labeling convention for the $F$-symbol. We define the shaded circuit element to associate the qudit state with an $F$-symbol phase. This phase is computed based on the $F$-symbol labels assigned in the circuit diagram (see also the discussion in Appendix~\ref{append:explicitcircuit}). (b) A circuit realization for moving the abelian anyons by extending the string operator (dark grey).}
    \label{appendix:fig:ab_circuit}
\end{figure}

The abelian anyon at the end of the string can be moved using a two-qudit unitary that satisfies
\begin{equation}
    \includegraphics[scale = 1]{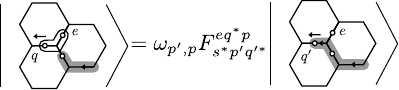},
     \label{moveabelian}
\end{equation}
again here 
$p = e^*\times q$, $p' = s\times p$ and $q' = s\times q$. We also define $\omega_{p',p}$ to be the same as $\omega_k$ in Eq.~\eqref{appendix:eq:omega} by replacing $i'_k = p'$ and $i_k = p$.
A schematic diagram of the unitary is shown in Fig.~\ref{appendix:fig:ab_circuit}. We give an example of the circuit realization for Eq.~\eqref{prepabelian} and Eq.~\eqref{moveabelian} in Appendix~\ref{append:explicitcircuit}.

{\bf Non-abelian quasiparticle strings:}
By placing one additional qudit at each end of the string operator, we can prepare a length-two non-abelian open string (corresponds to type-$s$ string) similar to the abelian case
\begin{equation}
    \includegraphics[scale = 1]{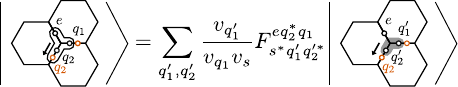}.
    \label{prepnonabelian}
\end{equation}
This four-qudit unitary uses two ancillary qudits to create a non-abelian string with two qudits along the path. The orange color highlights the qudits that store the ancillary qudit states. 

We can further define local unitary to move an existing non-abelian anyon at the endpoints of an open string. A key difference between the non-abelian and the abelian cases is that unitarily encoding non-abelian anyons with open string operators requires an additional ancillary qudit at the endpoint. In order to implement non-abelian strings, we need to ensure the endpoint of the string opeartor always lands on an edge that has an ancillary qudit. This can be achieved by placing one additional qudit on each edge. Alternatively, we can adapt the following strategy that only requires two additional qudits for implementing one open string operator.

\begin{figure}
    \centering
    \includegraphics[scale = 0.75]{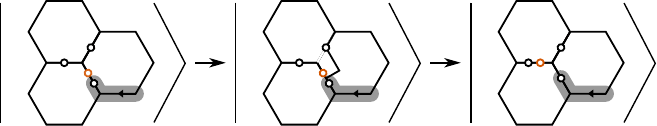}
    \caption{A local change of registered connectivity---to make sure the endpoint of the string can land at an edge with two qudits. No physical gates are needed here.}
    \label{appendix:fig:lattice_shift}
\end{figure}

\begin{figure}[t]
    \centering
    \includegraphics{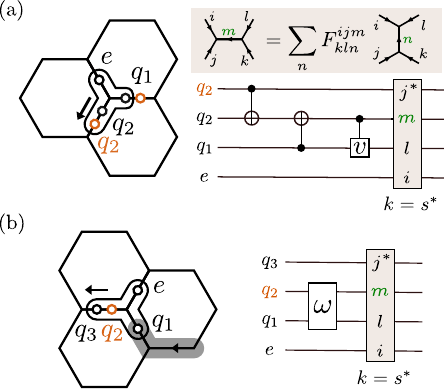}
    \caption{The quantum circuits for the non-abelian open string operators. The orange color highlights the qudits that store the ancillary qudit states. (a) A circuit realization for preparing non-abelian anyons. The shaded box shows the $F$-move defined by the $F$-symbol in the non-abelian model, together with the labeling convention for the $F$-symbol. The shaded circuit elements are qudit quantum gates that perform the $F$-move. The gates are controlled single-qudit rotations that change the state of the green qudit labeled with $m$. The gates are controlled by the states of the other input qudits, each of which is assigned with a label in the $F$-symbol according to the circuit diagram. (b) A circuit realization for moving the non-abelian anyons by extending the open string operator (dark grey).}
    \label{appendix:fig:nonab_circuit}
\end{figure}

For a given configuration, we first perform a local lattice distortion as shown in Fig.~\ref{appendix:fig:lattice_shift}. Such distortion does not require any physical quantum operation, one can simply register such change, e.g. on a classical computer. This step makes sure the endpoint of the string will land at an edge with two qudits. Following this virtual lattice distortion, we implement a  four-qudit unitary that satisfies
\begin{equation}
    \includegraphics[scale = 1]{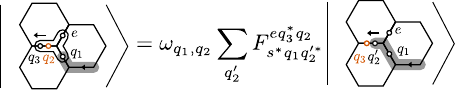}.
    \label{movenonabelian}
\end{equation}
The operators for creating and moving the anyons are schematically shown in Fig.~\ref{appendix:fig:nonab_circuit}. with a circuit in terms of qudit quantum gates. Notice that a long abelian string costs a constant depth quantum circuit while a linear-depth is required for the non-abelian strings.

\section{Isometry Property of Plaquette Operator $\hat{B}_p^s$ and the Open String Operator}\label{append:isometry}
Here we prove that $\hat{B}_p^s$ is an isoemtry on the given subspace as in Fig.~\ref{general_sn}a, when the representative qudit is in the null state initially. This follows directly from the algebraic definition of $\hat{B}_p^s$ in Eq.~\eqref{b_define}. We show this isometry property for the case when $h = 0$ (by continuity of the strings $b = g$ and $ c = i^*$), i.e. we compute the overlap
\begin{equation}
    \includegraphics[scale  =1]{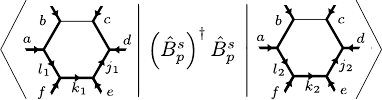}.
\end{equation}
Taking the algebraic definition of $\hat{B}^s_p$
\onecolumngrid
\begin{align*}
    &\sum_{g'i'j'k'l'} F^{bb^*0}_{s^*sg'^*}\left(F^{bb^*0}_{s^*sg'^*}\right)^*
    F^{c0c^*}_{s^*i's^*}\left(F^{c0c^*}_{s^*i's^*}\right)^*
    F^{dcj_1}_{s^*j'i'^*}\left(F^{dcj_2}_{s^*j'i'^*}\right)^*
    F^{ej_1^*k_1}_{s^*k'j'^*}\left(F^{ej_2^*k_2}_{s^*k'j'^*}\right)^*
    F^{fk_1^*l_1}_{s^*l'k'^*} \left(F^{fk_2^*l_2}_{s^*l'k'^*}\right)^*
    F^{al_1^*b}_{s^*g'l'^*}\left(F^{al_2^*b}_{s^*g'l'^*}\right)^* \\
     = & \sum_{g'i'j'k'l'}\delta_{s^*i'c} F^{bb^*0}_{s^*sg'^*}\left(F^{bb^*0}_{s^*sg'^*}\right)^*
    F^{dcj_1}_{s^*j'i'^*}\left(F^{dcj_2}_{s^*j'i'^*}\right)^*
    F^{ej_1^*k_1}_{s^*k'j'^*}\left(F^{ej_2^*k_2}_{s^*k'j'^*}\right)^*
    F^{fk_1^*l_1}_{s^*l'k'^*} \left(F^{fk_2^*l_2}_{s^*l'k'^*}\right)^*
    F^{al_1^*b}_{s^*g'l'^*}\left(F^{al_2^*b}_{s^*g'l'^*}\right)^* \\
     = &\delta_{j_1j_2}\delta_{k_1k_2}\delta_{l_1l_2}
     \sum_{g'}\delta_{s^*g'b^*}
    F^{bb^*0}_{s^*sg'^*}\left(F^{bb^*0}_{s^*sg'^*}\right)^*
    = \delta_{j_1j_2}\delta_{k_1k_2}\delta_{l_1l_2},
    \numberthis \label{b_isometry}
\end{align*}\twocolumngrid
\noindent
where we repeatedly used the unitarity of the $F$-symbol in Eq.~\eqref{unitarity}. The isometry property required in Eq.~\eqref{c_B} for C-$\hat B_p$ to be well-defined follows by setting $b = c = 0$. However, the isometry property holds even for $b,c\neq 0$, this allows us to define C-$\hat B_p$ on a more general subspace, it follows that any permissible order, other than the row-wise algorithm presented in the main texts, can be used for the string-net state construction.

In Appendix~\ref{appendix:string_op}, we claim there exist unitary operators that can be used to prepare and move the anyons on the string-net ground state. To see the conditions are compatible with unitarity, it suffices to consider the non-abelian case of the operators, the form of the abelian open string operator is a special case of the non-abelian string operator (see Appendix~\ref{appendix:string_op}). The isometry property of the operator for moving directly comes from the unitarity \eqref{unitarity} of $F$-symbol and $\omega$ being a phase. The isometry condition on the preparation operator in Eq.~\eqref{prepnonabelian} can be computed as
\begin{align*}
     &\bra{m p_1p_2\textcolor{orange}{p_2}}\left(\hat{U}^s_{\text{prep}}\right)^{\dagger}\hat{U}^s_{\text{prep}}\ket{e q_1q_2\textcolor{orange}{q_2}} 
     \\
     &= \sum_{q'_1,q'_2,p'_1,p'_2}\frac{d_{q'_1}}{d_{q_1}d_s} \left(F^{mp_2^*p_{1}}_{s^*q'_{1}q'^*_2}\right)^*F^{eq_2^*q_{1}}_{s^*q'_{1}q'^*_2}\delta_{me}\delta_{p_2q_2}\delta_{p'_1q'_1}\delta_{p'_2q'_2}
     \\
     &= \delta_{me}\delta_{p_2q_2}\delta_{q_1p_1}\sum_{q'_1}\frac{d_{q'_1}}{d_{q_1}d_s}\delta_{sq'^*_1q_1} 
     \\
     &=  \delta_{me}\delta_{p_2q_2}\delta_{q_1p_1},
     \numberthis\label{append:eq:isometry_op_prep}
\end{align*}
where the labels $m,p_1,p_2$ correspond to a qudit configuration matching $e,q_1,q_2$ in Eq.~\eqref{prepnonabelian}. In the second line of the calculation, we used Eq.~\eqref{unitarity} and the definition $d = |v|^2$. The last equality follows from Eq.~\eqref{append:eq:fusion}. 

Note that in Eq.~\eqref{append:eq:isometry_op_prep} it is important to have an ancillary qudit at the endpoint to orthogonalize the states. We expect that a similar construction using ancillary qudits can be used to realize non-simple open string operators. However, non-simple string operators will no longer correspond to a single string type, instead it is a mixture of several string types. Correspondingly, the phase factors $\omega_k$ in Eq.~\eqref{omega} associated with the strings are replaced with tensors~\cite{levin:2005}. The construction will likely have to be modified.

\begin{figure}
    \centering
    \includegraphics{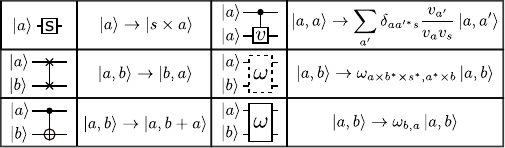}
    \caption{Definition for the qudit gates used in the construction of the open string operators, where $s$ is the label for the corresponding simple string operator. The quantity $\omega_{p',p}$ is the phase $\omega_k$ in Eq.~\eqref{appendix:eq:omega} with $i'_k = p'$ and $i_k = p$, and the orientation is to be identified from the direction of the string operators in the circuit diagrams, such as in Fig.~\ref{appendix:fig:ab_circuit} and~\ref{appendix:fig:nonab_circuit}.}
    \label{appendix:fig:gate_table}
\end{figure}
\section{Determine the $S$-matrix from Measurement}\label{append:smatrixfrommea}
The measurement of $S$-matrix relies on measuring the expectation of the unitary paths by an interferometry-like experiment (see section \ref{anyonbraiding}). The $S$-matrix is obtained by $S_{ab} = d_ad_bM_{ab}/\mathcal{D}$, where $M_{ab}$ is called the monodromy and is measured by braiding anyons $a$ and $b$, as shown in Eq.~\eqref{eq:braiding}. Sometimes we know the quantum dimension $d$ of the quasiparticles beforehand and sometimes we do not. However, tt turns out we can directly obtain the $S$-matrix just from the monodromy $M_{ab}$ without knowing the quantum dimensions. This is because we restrict ourselves to anyon models that are described by a unitary modular tensor category, where $S$-matrix is unitary. The constraint of unitarity implies $S^{\dag} = S^{-1}$, it follows that
\begin{equation}\label{append:eq:dmonodromy}
    \left(\frac{d_ad_b}{\mathcal{D}}\right)^2 = \frac{(M^{-1})^*_{ba}}{M_{ab}},
\end{equation}
where $M^{-1}$ is the inverse of the matrix $M_{ab}$. We can substitute this relation into Eq.~\eqref{eq:braiding}, $S$-matrix is directly given by 
\begin{equation}\label{append:eq:smonodromy}
    S_{ab} = \sqrt{\frac{(M^{-1})_{ba}^*}{M_{ab}}}M_{ab},
\end{equation}
solely in terms of the measurement outcomes. Due to the structure of the $S$-matrix, the monodromy $M_{ab}$ should also satisfy a set of constraints in order to yield a physical $S$-matrix. For example, from Eq.~\eqref{append:eq:dmonodromy} we can immediately conclude that $(M^{-1})^*_{ba}/M_{ab}>0$. By definition we have $S_{ab} = S_{ba}$, this suggests $M_{ab} = M_{ba}$. The list of constraints is not exhausted. Other constraints can come from, e.g. the Verlinde formula \cite{verlinde:1988} or being (together with the twist factors) the generator for the modular group \cite{bonderson:2006}. In practice, these stringent constraints and Eq.~\eqref{append:eq:smonodromy} will be very helpful in benchmarking the experimental data and identify the correct $S$-matrix for the underlying modular tensor category from the noisy data.

\begin{figure}
    \centering
    \includegraphics{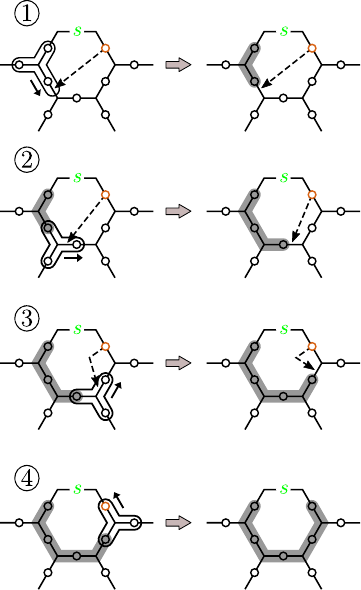}
    \caption{A circuit decomposition for C-$\hat B_p$ gate using the labeling convention in Eq.~\eqref{appendix:eq:cb_label}. All the gate operations are controlled by the qudit on the top bond (the green $s$). The dashed arrows show the virtual location of the ancillary qudit state, which is stored in the initially disentangled qudit (orange) throughout the steps. In step 1, the qudit gate is a qudit CNOT followed by Fig.~\ref{appendix:fig:nonab_circuit}a. The CNOT aligns the $\ket{0}$ qudit (orange) with the qudit at the endpoint of the string operator which is to be initialized. In step 2 and 3, the qudit gate is the same as Fig.~\ref{appendix:fig:nonab_circuit}b followed by a SWAP gate that keeps the ancillary qudit state in the orange qudit. The step 4 is similar to step 2 and 3, except that we adapt the circuit slightly since no ancillary qudit is required at the last step. The explicit circuit diagrams are given by Fig.~\ref{appendix:fig:explicit_gs_circ}.}
    \label{appendix:fig:gs_circuit}
\end{figure}

\section{Quantum Circuits for String Operators and C-$\hat B_p$}\label{append:explicitcircuit}
In this appendix, we provide some examples of quantum circuits for the string operators, which also lead to a quantum circuit for the C-$\hat B_p$. Since all the open string operators satisfy isometry condition, there are certain degrees of freedom in defining the unitary to realize them. What we show here is one realization of such unitary.

We encode the spin state on each site in a qudit. As a result, the quantum gates become qudit gates in general. For example, we can generalize the two-qubit CNOT gate to two-qudit CNOT by modular arithmetic, i.e. for $(N+1)$-level qudit, $\text{CNOT}_{\text{qudit}}\ket{a,b} = \ket{a,b+a}$ in mod $N+1$. Another example is the qudit SWAP gate that swaps the two states of two qudits, $\text{SWAP}_{\text{qudit}}\ket{a,b} = \ket{b,a}$. We will use the same circuit symbol for qudit CNOT and SWAP as for qubit CNOT and SWAP. For clarify, let us consider the case for abelian and non-abelian strings separately.

The abelian open string operators (i.e. each non-trivial value in the $F$-symbol is a phase), such as for the toric code, take a particularly simple form due to the fusion constraints (see Appendix~\ref{append:stringnetsummary}). A quantum circuit implementation of a type-$s$ string in terms of qudit gates are shown in Fig.~\ref{appendix:fig:ab_circuit}, where we define the colored circuit element in Fig.~\ref{appendix:fig:ab_circuit} as a two-qudit gate that associates each state with an $F$-symbol phase. Explicitly, they map $\ket{eq_1}\rightarrow F^{ejq_1}_{kln}\ket{eq_1}$ or $\ket{eq}\rightarrow F^{eq^*m}_{kln}\ket{eq}$ depending on which of two the labels $i,j,k,m,n,l$ in $F^{ijm}_{kln}$ are assigned to the input qudits according to the circuit diagram. The rest of the four unknown labels in the $F$-symbol are uniquely determined by fusion in the abelian model. The other qudit gates we used can be found in Table.~\ref{appendix:fig:gate_table}.
In the string preparation stage, we simplify the circuits by removing the phase in Eq.~\eqref{prepabelian} as discussed in Appendix~\ref{append:stringop}. The string preparation gate thus satisfies $\hat{U}^s_{\text{prep}}\ket{eq_1q_2} = F^{eq_2^*q_{1}}_{s^*q'_{1}q'^*_2}\ket{eq'_1q'_2}$, which is equivalent to Eq.~\eqref{prepabelian} up to an overall phase.

The non-abelian string circuits are shown in Fig.~\ref{appendix:fig:nonab_circuit}, where we have used the qudit CNOT introduced above. The colored gate in this case implements the $F$-move to the state according to the assigned labels in the circuit diagram. Unlike the abelian case, the gate changes the state of the qudit assigned with green label $m$ in Fig.~\ref{appendix:fig:nonab_circuit}. The definition for other qudit gates can be found in Table.~\ref{appendix:fig:gate_table}.

Next, we use the circuits for the string opeartors to construct a circuit for the C-$\hat B_p$. The controlled gate C-$\hat B_p$ can be seen as applying a type-$s$ string operator around the plaquette if the control qudit of the gate (i.e. the representative qudit for the plaquette) is in state $\ket{s}$. It thus follows that we can decompose C-$\hat B_p$ into a sequence of local controlled gates that prepare and deform the string operator as in Fig.~\ref{appendix:fig:ab_circuit} and Fig.~\ref{appendix:fig:nonab_circuit}. We show how this can be done for the non-abelian case, the abelian case follows straightforwardly. Note that the subspace C-$\hat B_p$ acts on is
\begin{equation}\label{appendix:eq:cb_label}
    \includegraphics[scale = 1]{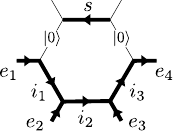}.
\end{equation}
Our goal is to apply a type-$s$ string operator around the plaquette, where $\ket{s}$ is the state of the control qudit at the top. 

\begin{figure}
    \centering
    \includegraphics{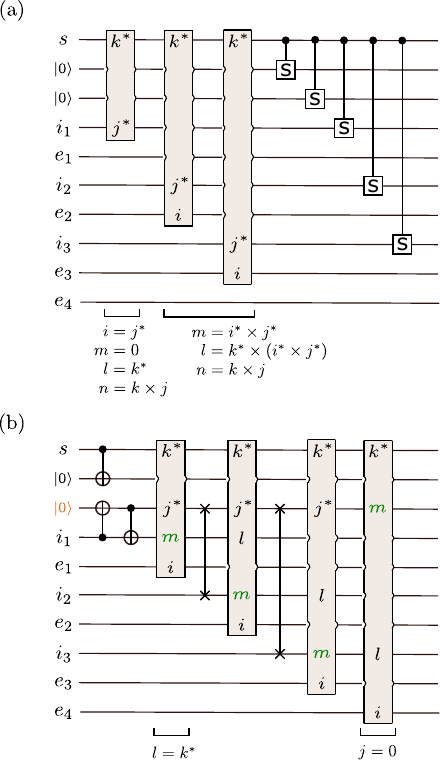}
    \caption{Explicit circuit diagrams for Fig.~\ref{appendix:fig:gs_circuit}. The qudits are labeled by Eq.~\eqref{appendix:eq:cb_label}. (a) The C-$\hat B_p$ for the abelian case. Here we have used the controlled fusion gate C-$S$, which performs the unitary C-$S\ket{s,i} = \ket{s,s\times i}$. (b) The circuit for the non-abelian case, the qudit used to store the ancillary state is marked orange.}
    \label{appendix:fig:explicit_gs_circ}
\end{figure}

A schematics of the circuit is shown in Fig.~\ref{appendix:fig:gs_circuit}. To construct C-$\hat B_p$, we first initialize a length-two string using the circuits in Fig.~\ref{appendix:fig:nonab_circuit}a. In this step, we use one of the initially disentangled qudits, which is chosen to be the one on the right in this example, to store the ancillary qudit state. Next, we extend the string around the plaquette by applying the circuit in Fig.~\ref{appendix:fig:nonab_circuit}b followed by a qudit SWAP gate that ensures the ancillary qudit state is always stored in the same qudit throughout. In the last step, the endpoint of the string is going to land at the qudit that stores the ancillary qudit state, We do this with a slight adaptation of the circuit by fixing the state $q_3$ to $\ket{0}$ in Fig.~\ref{appendix:fig:nonab_circuit}b. In other words, since the ancillary qudit is supposed to store the initial state of the endpoint qudit before the string operator is applied, the use of the ancillary qudits becomes unnecessary if that initial qudit is disentangled $\ket{0}$. The resulting explicit circuit diagrams for the decomposition are depicted in Fig.~\ref{appendix:fig:explicit_gs_circ} for both abelian and non-abelian cases. One can check that for the case $s = 0$, the circuit operates an identity on the input state.

We note that all the qudit gates that implement the corresponding $F$-symbol can be realized as a multi-controlled qudit gate. The gate applies a single-qudit rotation on the qudit marked by a green label depending on the states of the other qudits. We illustrate this in the next section with an explicit example.

\section{The circuits for double Fibonacci Model}\label{append:dfibexample}
\begin{figure}[t]
    \centering
    \includegraphics{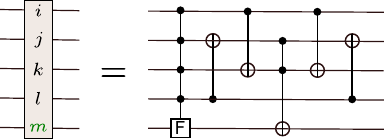}
    \caption{A quantum circuit for the $F$-symbol given in Ref~\cite{bonesteel:2012}, where all the gates are qubit gates. Here we have used a four-qubit controlled rotation gate, i.e. a single-qubit rotation is applied to the fifth qubit if the qubits at the solid dots are $\ket{1}$, otherwise no operation will be performed. The single-qubit rotation is defined in Eq.~\eqref{appendix:eq:f_sym}.}
    \label{appendix:fig:dfib_f_circ}
\end{figure}
The double Fibonacci model is a string-net model that hosts a non-abelian topological order. The model can be defined on a honeycomb lattice with qubit ($\ket{0}$ or $\ket{1}$ correspond to string-0 and string-1) on each site. The strings are unoriented with branching rules given by $\delta_{111} = \delta_{110} = \delta_{000}=1$. The $F$-symbol is defined by the Fibonacci unitary tensor category
\begin{alignat}{2}
    v_0 &= 1, \quad v_1 = \phi^{\frac{1}{2}}, \nonumber \\
    F^{111}_{110} &= \phi^{-\frac{1}{2}},\quad F^{111}_{111} & &= -\phi^{-1}, \nonumber\\
    F^{110}_{110} &= \phi^{-1}, \quad F^{110}_{111} & &= \phi^{-\frac{1}{2}} , \label{appendix:eq:f_sym}
\end{alignat}
where $\phi = \frac{1+\sqrt{5}}{2}$ is the golden ratio \cite{levin:2005}. All the other non-zero values in $F$-symbol are 1.

As mentioned the last section, a circuit implementation of the $F$-move can be realized by a multi-controlled qubit gate. For the double Fibonacci model this gate can be decomposed as shown in Fig.~\ref{appendix:fig:dfib_f_circ}~\cite{bonesteel:2012}, where the single-qubit rotation is given by the $F$-symbol in Eq.~\eqref{appendix:eq:f_sym} as $(F)_{ij} = F^{11i}_{11j}$.

We can insert the circuit of the $F$-move into Fig.~\ref{appendix:fig:nonab_circuit} and \ref{appendix:fig:explicit_gs_circ}b to obtain a set of circuits that efficiently simulate\leotp{s} the double Fibonacci model on a digital quantum computer. Although the circuit has not yet been  optimized for practical implementation (this depends on the connectivity of the device), we can still roughly estimate the resources needed for the simulation by counting the number of CNOT gates needed to do a C-$\hat B_p$ and perform an anyon movement. The multi-controlled $F$ gate in Fig.~\ref{appendix:fig:dfib_f_circ} can be broken down into 30 CNOT gates \cite{barenco:1995}. The Toffoli gate can be further decomposed into 6 CNOT gates~\cite{shende:2009}, together with the other 4 CNOT we have 40 CNOT gates in total. For the C-$\hat B_p$ operator in Fig.~\ref{appendix:fig:explicit_gs_circ}b, the first and last $F$-move unitary can be simplified exploiting the fact that we are not using the full $F$-symbol. In the first $F$-move, the four-qubit controlled gate in Fig.~\ref{appendix:fig:dfib_f_circ} becomes a three-qubit controlled gate by identifying $l=k$, which can be decomposed into 13 CNOT \cite{barenco:1995}, altogether the first $F$-move consists of 23 CNOT. Similarly for the last $F$-move, by setting $j = 0$, the circuit in Fig.~\ref{appendix:fig:dfib_f_circ} can be simplified to 8 CNOT. Taking into account these consideration, there are in total 120 CNOT gates (without any circuit optimization).

The counting above does not take into account any circuit parallelization. In practice, many of the gates can be operated in parallel to improve the efficiency, and circuit simplification may be exploited. For example, the continuity of the strings on the subspace in Eq.~\ref{appendix:eq:cb_label} implies $e_1=i_1$ and $i_3=e_4$, this can be used to further reduce the number of gates needed in Fig.~\ref{appendix:fig:explicit_gs_circ}.

\end{document}